\documentclass[a4paper,twocolumn,prb,amssymb,aps]{revtex4}
\usepackage{dcolumn}
\usepackage{bm}
\usepackage{xspace}

\usepackage{graphicx}
\usepackage{amssymb}

\newcommand{\etal}{{\it et al.}}
\newcommand{\scl}{$\mathrm{(Sr,Ca,La)_{14}Cu_{24}O_{41}}$}
\newcommand{\sr}{$\mathrm{Sr_{14}Cu_{24}O_{41}}$}
\newcommand{\srca}{$\mathrm{Sr_{14-x}Ca_xCu_{24}O_{41}}$}
\newcommand{\lav}{$\mathrm{La_{4}Sr_{10}Cu_{24}O_{41}}$}
\newcommand{\lacasr}{$\mathrm{La_{x}(Ca,Sr)_{14-x}Cu_{24}O_{41}}$}

\newcommand{\figref}[1]{Fig.~\protect\ref{#1}}

\begin{document}




\title{Magnetization of undoped 2-leg $S=\frac{1}{2}$ spin ladders in La$_{4}$Sr$_{10}$Cu$_{24}$O$_{41}$}


\author{R. Klingeler, I. Hellmann, P. Ribeiro, C. Hess, B. B{\"u}chner}

\affiliation{Leibniz-Institute for Solid State and Materials Research IFW Dresden, 01171 Dresden,
Germany}

\begin{abstract}
Magnetization data of single crystalline \lav\ are presented. In this compound, doped spin chains
and undoped spin ladders are realized. The magnetization, at low temperatures, is governed by the
chain subsystem with a finite interchain coupling which leads to short range antiferromagnetic
spin correlations. At higher temperatures, the response of the chains can be estimated in terms of
a Curie-Weiss law. For the ladders, we apply the low-temperature approximation for a $S=1/2$ 2-leg
spin ladder by Troyer et al.
\end{abstract}



\maketitle \narrowtext

The compounds \scl\ possess an incommensurate layered structure of two alternating
subsystems~\cite{McCarron88}. In these subsystems, two quasi one-dimensional (1D) magnetic
structures are realized which are oriented along the $c$-axis, i.e. $S=\frac{1}{2}$ Cu$_2$O$_3$
spin ladders and CuO$_2$ spin chains. Since the spin ladders exhibit a considerable spin gap
$\Delta$ of several hundred Kelvin, the low temperature magnetic response of these compounds is
described in terms of the weakly coupled CuO$_2$ spin chains~\cite{Klingeler06}. In addition, the
compounds \srca\ are intrinsically hole doped with six holes per formula unit. $\mathrm{La^{3+}}$
doping leads to an decrease of the hole content in both the ladder and the chain subsystem. For
\lacasr , with $x\geq 3$, the holes completely reside at chain sites~\cite{Nuecker00}. The
magnetic coupling of Cu-spins via a hole is antiferromagnetic (AF) while a ferromagnetic (FM)
superexchange is realized for adjacent Cu spins. Interchain coupling leads to a 3D AF spin
ordering for $x\geq 5$, below $\sim$10\,K.

\begin{figure}[ht!]
\center{\includegraphics*[width=\columnwidth]{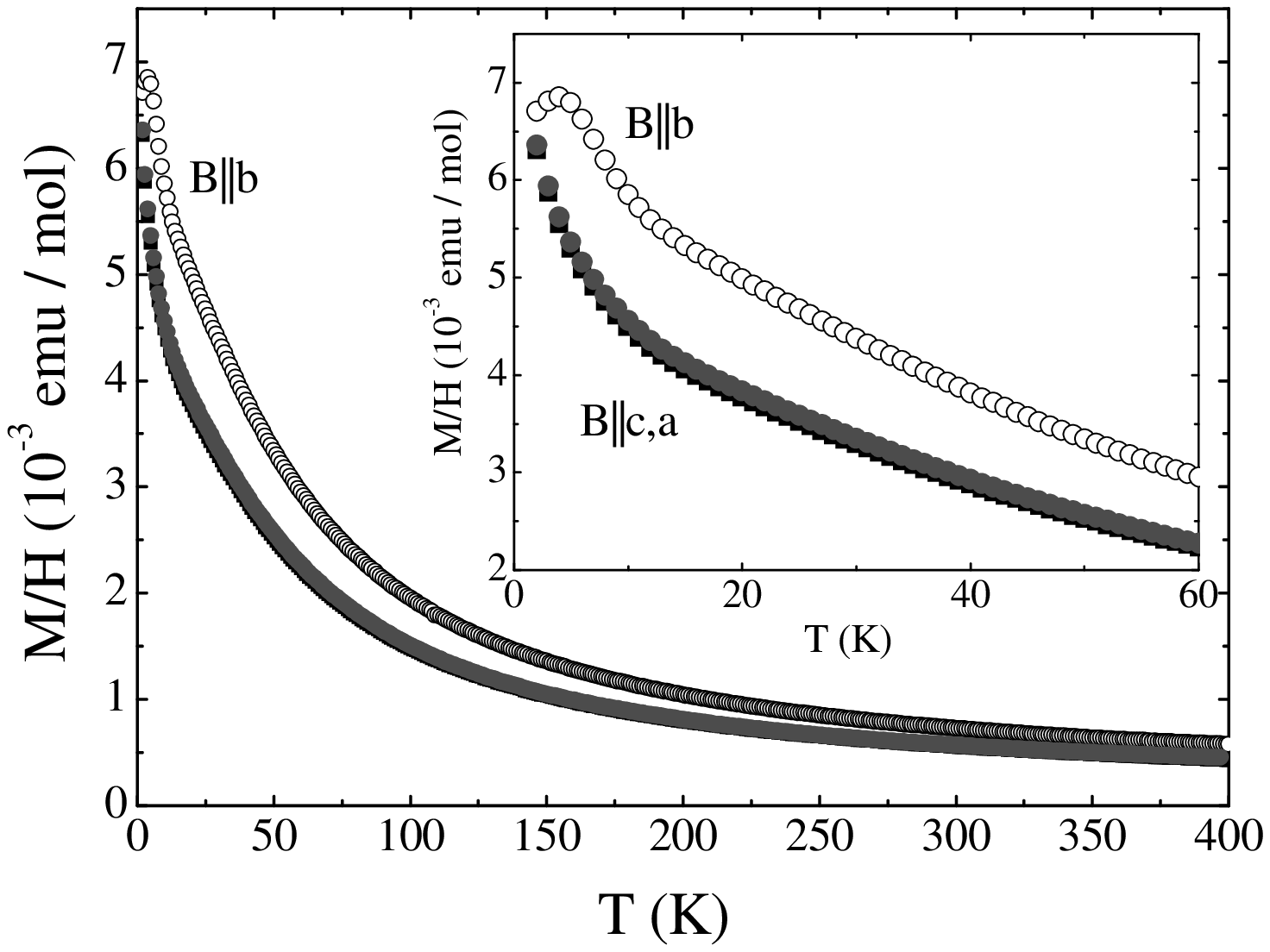}}
  \caption{Static Susceptibility $M/H$, with $\mu_0H=1$\,T applied along the $a$, $b$, and $c$-axis, respectively.
  The inset highlight the low temperature regime.} \label{fig1}
\end{figure}

The magnetization data in \figref{fig1} mainly reflect the chain magnetism. At high temperatures,
i.e. at 400\,K, the anisotropy is well explained by the anisotropic $g$-factor ($g_\perp\approx
2.35$ and $g_\|\approx 2.05$) and the anisotropy of the Van-Vleck magnetism. These extracted
values for $g$ agree well with recent ESR data~\cite{Kataev}. Upon cooling, an approximate
Curie-Weiss behavior is observed, with an additional upturn of $M$ below $\sim$10\,K and a maximum
for $B\|b$ at 4\,K. The exact origin of the low temperature features of the magnetization is
unknown but we recall that, in \lav , along the chains there are several competing magnetic
interactions, i.e. a FM nearest neighbor coupling, an AF next-nearest neighbor coupling and an AF
coupling via the $\sim$20\% of holes. In addition, there is an AF interchain coupling, which
yields strong AF spin correlations with an easy axis parallel to the $b$-axis as it is reflected
by the data. Long range AF spin order, however, does not evolve above 2\,K.

\begin{figure}[ht!]
\center{\includegraphics*[width=\columnwidth]{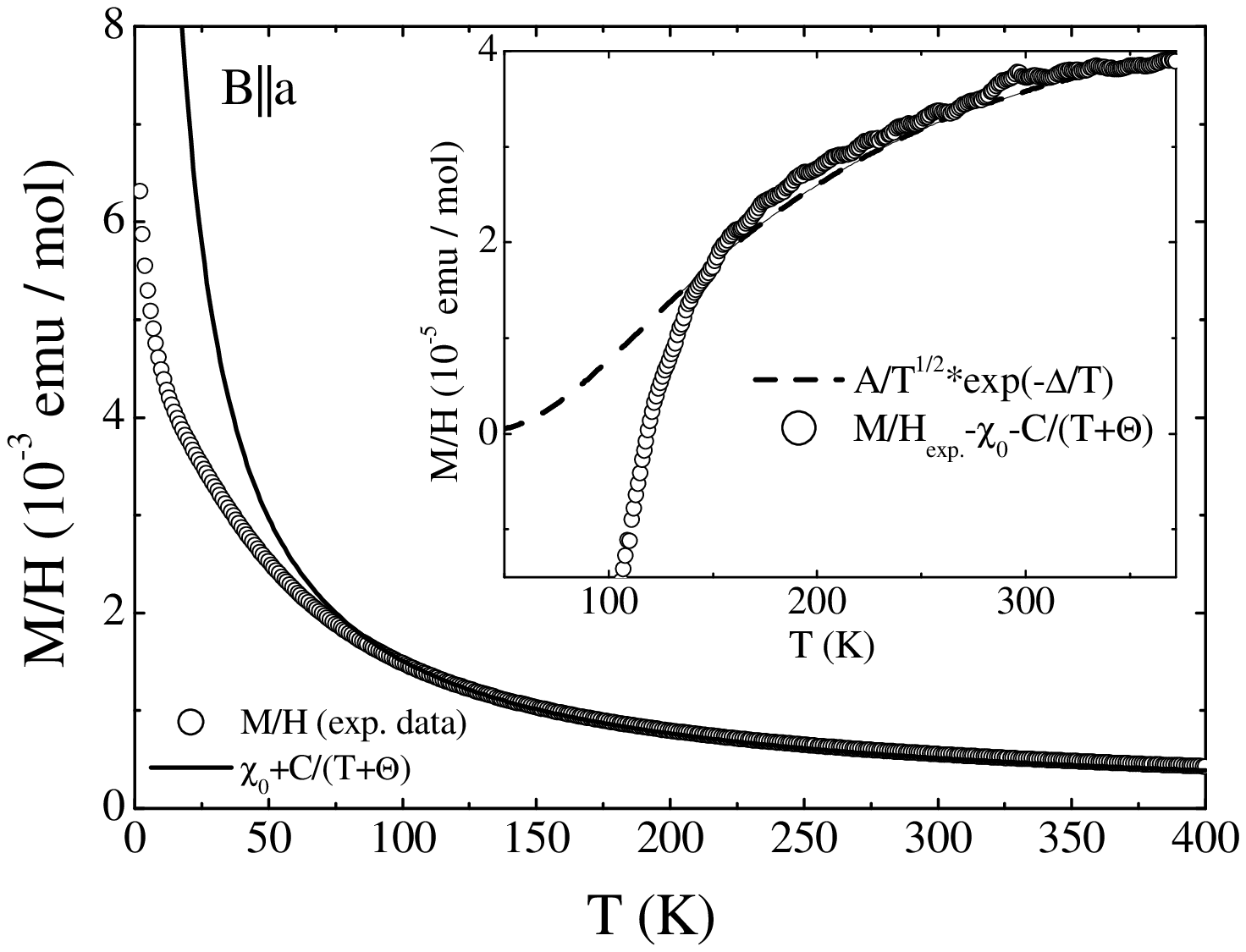}}
  \caption{Static susceptibility $M/H$ (data points) and magnetic response of the chains (dashed line) as
  described by Eq.~\ref{eins} for high temperatures.
  Inset: Curie-Weiss-like susceptibility subtracted from the experimental data (data points) and magnetic susceptibility of isolated 2-leg ladders
  (dashed line).}\label{fig2}
\end{figure}

At high temperatures, i.e. $T\gtrsim 140$\,K, the magnetization of the chains can be described in
terms of the Curie-Weiss law
\begin{equation}
\chi(T)=\chi_0+\frac{C}{T+\Theta}.\label{eins} \end{equation} \noindent In addition to the chain
magnetism, the magnetic response of the ladders has to be considered at higher temperatures. The
spin gap of the ladders leads to an exponential increase of the $\chi_{\rm ladders}$. Within the
low-temperature approximation for a $S=1/2$ 2-leg spin ladder, the ladder contribution to the
susceptibility $\chi_l$ may be described by

\begin{equation}
\chi_{l}(T)= \frac{A}{\sqrt{T}}e^{-\Delta /(k_{\rm B}T)},\label{troyer}
\end{equation}

\noindent with $\Delta$ being the spin gap~\cite{troyer}. For the analysis of the magnetization
data, we hence fitted our data by the sum of Eqs.~\ref{eins} and \ref{troyer}. We fixed the spin
gap to the value of $\Delta \simeq 313$\,K known from neutron scattering on the same
compound~\cite{Notbohm}. We restricted the analysis to temperatures above 140\,K. The results of
this procedure, for $B\|a$, are displayed in \figref{fig2}. Apparently, the data are reasonably
well described by Eqs.\,1 and 2 at $T\gtrsim 140$\,K. At low temperatures below 140\,K, however,
there are strong deviations which we attribute to the fact that, in this temperature regime, due
to the intrachain couplings the chain response can not be described by a Curie-Weiss law.
Quantitatively, the analysis yields
reasonable results for the chain magnetism, i.e., $C_a=(0.152\pm 0.1)$\,erg\,K/(G$^2$mol Cu),
$C_b=(0.196\pm 0.1)$\,erg\,K/(G$^2$mol Cu), and $\Theta\approx 1$\,K. For the ladders, we obtain:
$$A_{a,c}=(1.7\pm 0.1)\cdot 10^{-3}\,{\rm erg}\,\sqrt{\rm K}/({\rm G}^2{\rm mol Cu}),$$
$$A_b=(1.46\pm 0.1)\cdot 10^{-3}\,{\rm erg}\,\sqrt{\rm K}/({\rm G}^2{\rm mol Cu}).$$

The value of the parameter $A$ changes significantly if the chemically undoped compound \sr\ is
considered. Here, the spin gap of the ladders amounts to $\Delta=377$\,K.~\cite{Eccleston}
Analyzing the data presented in Ref.~\cite{Klingeler06} in the same way as described above yields
$$A_c=(4.1\pm 0.1)\cdot 10^{-3}\,{\rm erg}\,\sqrt{\rm K}/({\rm G}^2{\rm mol Cu}).$$ We mention
that, in \sr , the fitting has to be restricted to $T>200$\,K since there is a charge ordering
transition below that temperature\cite{Hess}. Moreover, in \sr\ the ladders contain a considerable
amount of holes while there are no holes in the ladders of \lav .

In summary, we have presented the magnetization of the spin chain and spin ladder compound \lav .
The low temperature response is governed by the chain subsystem in which interchain couplings lead
significant AF correlations. At high temperatures the chain magnetism follows the Curie-Weiss law.
Additional contributions to the magnetization are due to the undoped spin ladders.


\begin{thebibliography}{00}

\bibitem{McCarron88}
E. McCarron \etal , Mat. Res. Bull. \textbf{23}, 1355 (1988)

\bibitem{Nuecker00}
N. N{\"u}cker \etal , Phys. Rev. B \textbf{62}, 14384 (2000)

\bibitem{Kataev}
V. Kataev \etal , Phys. Rev. Lett. \textbf{86}, 2882 (2001)

\bibitem{Klingeler05b}
R. Klingeler \etal., Phys. Rev. B \textbf{72}, 184406 (2005)

\bibitem{Klingeler06}
R. Klingeler \etal , Phys. Rev. B \textbf{73}, 014426 (2006)


\bibitem{Troyer94}
M. Troyer, H. Tsunetsugu, D. W{\"u}rtz, Phys. Rev. B \textbf{50}, 13515 (1994)

\bibitem{Notbohm}
S. Notbohm \etal , to be pubished.

\bibitem{Eccleston}
R.S. Eccleston \etal , Phys. Rev. Lett. \textbf{81}, 1702 (1998)

\bibitem{Hess}
C. Hess \etal , Phys. Rev. Lett. \textbf{93}, 027005 (2004)


\end{thebibliography}
\end{document}